\documentclass[conference]{IEEEtran}
\IEEEoverridecommandlockouts
\usepackage{cite}
\usepackage{amsmath,amssymb,amsfonts}
\usepackage{algorithmic}
\usepackage{graphicx}
\usepackage{textcomp}
\usepackage{xcolor}
\def\BibTeX{{\rm B\kern-.05em{\sc i\kern-.025em b}\kern-.08em
    T\kern-.1667em\lower.7ex\hbox{E}\kern-.125emX}}
    
\usepackage{nicefrac}
\usepackage{siunitx}
\usepackage{array,framed}
\usepackage{booktabs}
\usepackage{
  float,
  epsfig,
  wrapfig,
  graphics,
  graphicx,
  subcaption
}    
\usepackage{graphics}
\usepackage{xparse} 
\usepackage{xspace}
\usepackage{multirow}
\usepackage{csvsimple}
\usepackage{balance}
\usepackage{listings}    
\usepackage{colortbl}
\usepackage{
  tikz,
  pgfplots,
  pgfplotstable
}
\usepackage{hyperref}   

\usetikzlibrary{
  shapes.geometric,
  arrows,
  external,
  pgfplots.groupplots,
  matrix
}

\pgfplotsset{compat=1.9}

\definecolor{codegreen}{rgb}{0,0.6,0}
\definecolor{codegray}{rgb}{0.5,0.5,0.5}
\definecolor{codepurple}{rgb}{0.58,0,0.82}
\definecolor{backcolour}{rgb}{0.95,0.95,0.92}

\lstdefinestyle{mystyle}{
  backgroundcolor=\color{backcolour}, commentstyle=\color{codegreen},
  keywordstyle=\color{magenta},
  numberstyle=\tiny\color{codegray},
  stringstyle=\color{codepurple},
  basicstyle=\ttfamily\footnotesize,
  breakatwhitespace=false,         
  breaklines=true,                 
  captionpos=b,                    
  keepspaces=true,                 
  numbers=left,                    
  numbersep=5pt,                  
  showspaces=false,                
  showstringspaces=false,
  showtabs=false,                  
  tabsize=2
}

\begin{document}

\title{A Tool For Debugging Quantum Circuits
}

\author{\IEEEauthorblockN{Sara Ayman Metwalli}
\IEEEauthorblockA{\textit{Keio Quantum Computing Center} \\
\textit{Keio University}\\
Fujisawa, Japan \\
sara@sfc.wide.ad.jp}
\and
\IEEEauthorblockN{Rodney Van Meter}
\IEEEauthorblockA{\textit{Keio Quantum Computing Center} \\
\textit{Keio University}\\
Fujisawa, Japan \\
rdv@sfc.wide.ad.jp}
}

\maketitle



\begin{abstract}
    As the scale of quantum programs grows to match that of classical software, the nascent field of quantum software engineering must mature and tools such as debuggers will become increasingly important. However, developing a quantum debugger is challenging due to the nature of a quantum computer; sneaking a peek at the value of a quantum state will cause either partial or complete collapse of the superposition and may destroy the necessary entanglement.  As a first step to developing a full quantum circuit debugger, we have designed and implemented a quantum circuit debugging tool. The tool allows the user to divide the circuit vertically or horizontally into smaller chunks known as \emph{slices}, and manage their simulation or execution for either interactive debugging or automated testing. The tool also enables developers to track gates within the overall circuit and each chunk to understand their behavior better. Feedback on usefulness and usability from early users shows that using the tool to slice and test their circuits has helped make the debugging process more time-efficient for them.
\end{abstract}



\section{Introduction}
\label{sec:intro}

\subsection{The Need For a Quantum Debugging Tool}
\label{intro}
The forty-year history of quantum computers has taken us through initial curiosity, naive optimism, then dismay at the scale of proposed error-corrected systems, and into today's excitement over the availability of real, but still small and error-prone, systems~\cite{preskill2018quantum,ladd10:_quantum_computers,van-meter19:_quant_tweet_zen}.  Algorithms have followed a similar roller coaster, arriving at the point where modest demonstration implementations of algorithms originally defined as abstract equations in theory papers are now common~\cite{montanaro2015:qualgo-qi}. The challenge on both hardware and software now is scalability: more qubits and larger, more sophisticated programs where, unlike today's demonstrations, the results are not \emph{a priori} known.  Working at large scale implies the need for a mature software engineering (SE) approach, including tools for all phases of the life cycle.

\begin{figure*}[htbp]
\centerline{\includegraphics[scale=0.25]{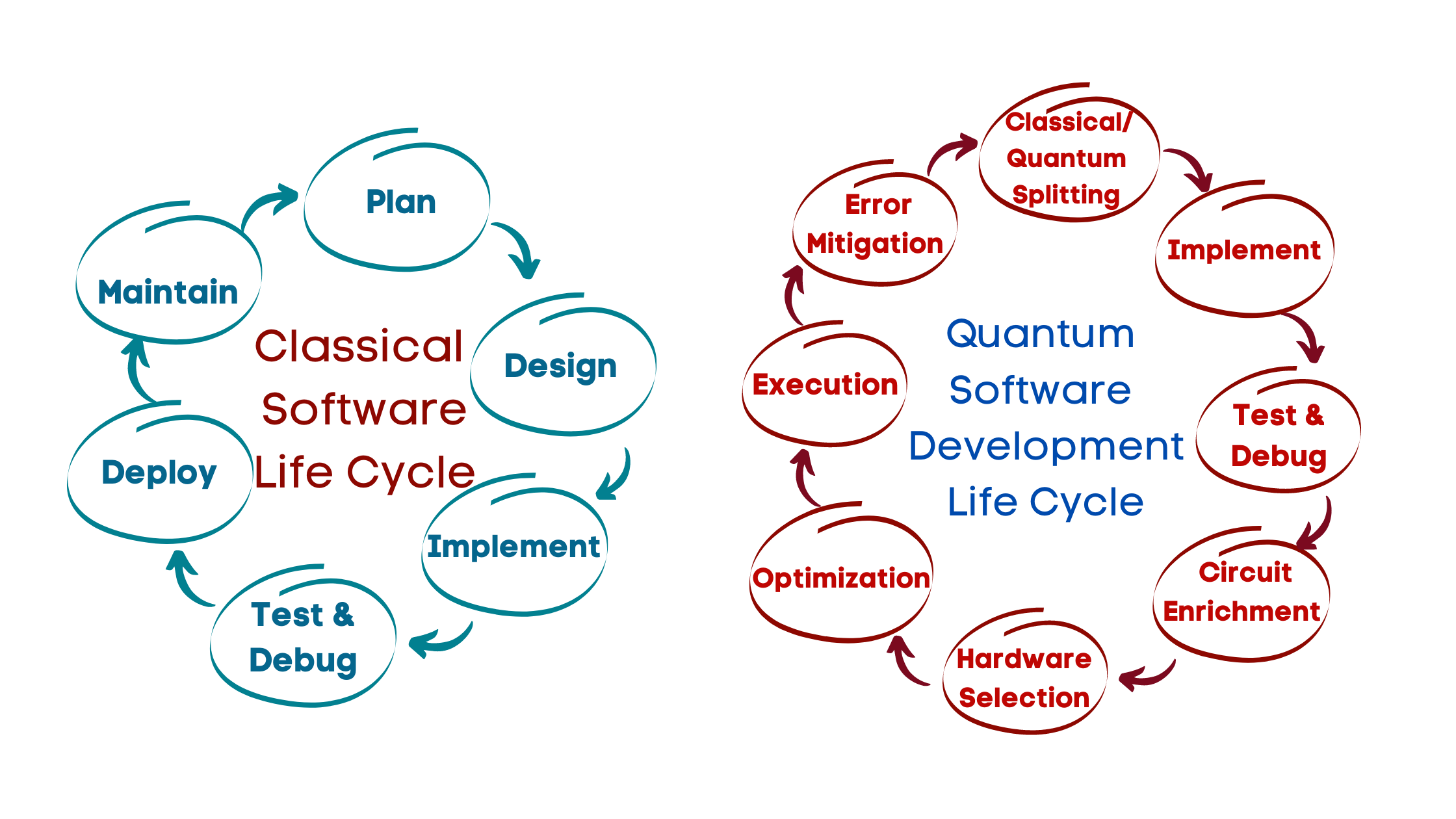}}
\caption{A general life cycle of classical software vs quantum software as described in~\cite{weder2020quantum}.}
\label{cycle}
\end{figure*}

Software engineering follows a certain cycle, and is a reasonably mature process. Two important phases are the design and development of the key conceptual elements, and testing and fixing bugs~\ref{cycle}. The conceptual elements may be supported with formal specifications, pseudocode, libraries, modeling tools and languages, etc. Bugs arise from mistakes in the specification of a program, or from mistakes in translating the specification into code (or, sometimes, from bugs in the tools themselves).  A variety of methods, both formal and informal, are used to find such bugs and to prevent their recurrence once isolated and fixed. Unit testing, regression testing and continuous integration, path coverage testing, and the many types of test cases that software testers construct all contribute to locating and eliminating the different types of bugs. Using these techniques and tools, it is now possible to construct and support systems as complex as tens of millions of lines of code, as in the Linux kernel and other similar systems.

If we consider the quantum software development cycle described in~\cite{weder2020quantum} as shown in Fig.~\ref{cycle}, we can find general similarities to the classical software development cycle.
There are two major differences in debugging quantum programs. Quantum computers can operate on \emph{superpositions} of values, each with a \emph{complex amplitude}~\cite{sutor19:dancing,preskill:PH-CS219}. It might be assumed that the exponential growth in the state space poses a fundamental problem, but in fact the problem is more nuanced. $n$ classical bits can carry any of $2^n$ values, yet we do not worry about testing all $2^n$ input values for a program, let alone all of the astronomical number of possible states when we include temporary variables. (After all, a system with 1GB of RAM has $2^{8,589,934,592}$ possible states, almost all of which will never be reached.) Instead, we focus on key paths and expected input values, and work to build robust error handling for the vast majority of unwanted states. With quantum, the process is different, and we often have to consider the behavior of all of the possible inputs.

The first, most important, challenge lies in the nature of quantum algorithms: their goal is not to simply find a solution to a problem, but instead to build \emph{interference patterns} that raise the amplitude of correct solutions to problems at the expense of incorrect solutions. Algorithms consist of blocks that perform essentially classical computations, and blocks that use the state so created to effect constructive and destructive interference. Thus, when dealing with quantum circuits, we can divide the behaviour of blocks (or \emph{slices}) into two main categories:

\begin{itemize}
    \item Pseudo-classical slices: These are circuits that imitates the behaviour of classical gates but perform on superposition, such as quantum adders and Grover's oracle.
    \item Full-quantum slices: These are circuits that alter the distribution of amplitudes, especially by creating interference patterns, such as the Grover diffusion operator and the quantum Fourier transform. 
\end{itemize}

Based on the different characteristics of those types, the process of debugging them will differ greatly. It will even differ within each type. If we look closely to the pseudo-classical circuits, we can further categorize it into simple and complex pseudo classical circuits. 
These two cases, which are essentially classical reversible logic~\cite{bennett73:reversible,bennett88:_notes}, are distinguished primarily by the difficulty of generating test cases with adequate coverage. For an adder, for example, we might test a few simple inputs, test overflow cases, and reason by induction about the rest. For a complex slice, we want to know if it correctly identifies a solution to the problem, but we do not \emph{a priori} know a solution. (If we did, we would have no need for the computer!) This is often addressed by building simple, small test cases, then extending to larger problems. In quantum systems, we can stick with similar techniques and basic unit testing and debugging, without the need to address the issue of exponentially large superpositions.

The challenge arises when we are addressing the full-quantum slices, because these circuits contain quantum properties we can't really debug using traditional debugging techniques. \emph{How do we debug the changing amplitudes and confirm the interference patterns we need?} Similar to the pseudo-classical circuits, we can further separate full-quantum circuits based on their size and complexity, into simple and complex circuits. Although creating test vectors for either type is not as simple as doing so for the pseudo-classical circuits, coming up with test vectors for simple full-quantum circuits should be more straightforward than the complex ones. For example, creating test vectors for a 4-qubit quantum circuit with only 4 Hadamard gates is simpler than creating test vectors for a circuit of 10-qubits and 50 different quantum gates.

Another major challenge is quantum circuits that are problem-specific. Rather than a data-independent loop structure, for example, input values are generally expressed as a series of gates. A graph state problem, for example, might be compiled to include gates corresponding to each edge in the graph~\cite{metwalli2020finding}. Thus, extending from the simple test cases to a specific problem can be an error-prone process, and our SE methods and tools must address this issue.

\subsection{The state of quantum software today}
In the future, we can expect that problem-specific compilation will become an automated process, but today the programmer specifies gates by hand, making validation an important step. However, supporting both newcomers to the field as well as experienced quantum programmers as they attempt to build intuition and learn to create the interference patterns that drive quantum algorithms will require building new tools and defining new concepts~\cite{ali2022}.

Today, there are approaches developers could take to transform their algorithms and ideas into quantum circuits~\cite{heim2020quantum,gay05:_quant_progr_lang}, and if they are small/ medium-scale, they can attempt to execute it on an actual quantum hardware. These available approaches differ based on the core programming model. We can generally divide these approaches into four categories.
\begin{itemize}
    \item High-level quantum programming language supporting the developer's quantum intuition and allowing them to use it to build and design quantum algorithms. Examples of such programming languages are Sliq~\cite{bichsel2020silq} and Quipper~\cite{green2013quipper}.  
    \item Gate-level programming. In this option, the developer translates his idea into a sequence of gates and then either simulates this circuit, visualizes it or runs on a hardware device. We can divide options within this category even further.
    \begin{itemize}
        \item Building the circuit using code, often using a classical-language-supported library or package. Examples of this option are Qiskit~\cite{aleksandrowicz2019qiskit}, Cirq~\cite{ai2018}, and PyQuil~\cite{pyq}.
        \item Using a drag-and-drop tool to build the circuit and then simulate the results and view them visually. These tools include, QUI~\cite{h2018}, the IBM Circuit Composer~\cite{santos2016ibm}, and Quirk~\cite{gidney}.
        \item Using the Quantum Assembly language, or QASM~\cite{cross2017open}.
    \end{itemize}
    \item Building the circuit using other compilation paradigms.
    \begin{itemize}
        \item Using a low-level approach, for example, using pulses and signals to directly control the quantum hardware, the main example for that is OpenPulse~\cite{aleksandrowicz2019qiskit}.
        \item Using a more quantum physics and mechanics approach like using the ZX-calculus~\cite{kissinger2019pyzx}. 
        \item Circuit optimizer, back-end compilers and interpreters. For example Tket~\cite{sivarajah2020t}, TriQ~\cite{murali2019full}, and Qbsolv~\cite{booth2020qci}.
    \end{itemize}
\end{itemize}

All of these tools, regardless of their compilation model, focus on the current generation of hardware, on small programs and on the important problems of optimization and mapping to specific processors~\cite{chong2017programming,Siraichi:2018:QA:3179541.3168822,nishio20:error-aware}, as well as tackling designing and implementing programs for hybrid, or adaptive, algorithms~\cite{farhi2014quantum,HOGG2000181,mcclean2016theory,peruzzo2014variational,Trugenberger_2002}. For now, most of these tools leave it to the programmer to mentally plan the algorithm and to examine the outputs to assess whether the program operates as intended, though some researchers are working to advance formal verification methods into the quantum arena~\cite{rand2018qwire}.\\
However, we need to start building tools for the future of quantum, because as we move toward large scale, key elements are missing in the current quantum software development tool-chain, including a quantum debugger and the tools for automating program testing such as unit tests. Both of these depend on the ability to isolate a portion of a quantum circuit, examine and understand its inner functionality. More specifically, the programmer needs to understand the interfaces among the parts, and to prepare input vectors and check their corresponding outputs without paying exponential costs in state or time spaces, whether the circuit is being simulated or executed on an a quantum computer. This, however, will not be possible as the circuit size increases 
and classical devices fail to simulate them. Inevitably, of course, working on a sub-circuit in this fashion raises questions about how to assess correctness \emph{without} the exponential number of terms in a quantum superposition. As quantum debugging becomes a focus and an essential skill for the current and next generations of quantum circuits and developers, the value of a tool that enables the understanding of the circuit as well as the error reasons will only increase.

Solving the challenge of debugging quantum circuits is not going to be a simple task, but it is definitely a challenge that we need to address and attempt to solve the best we can, and that is the target of this work.
In this paper, we present our quantum circuit debugging implemented and developed using Python and Qiskit . The tool will allow developers to divide their circuits into smaller chunks, categorize these chunks, and test them making the process of locating and fixing bugs relatively easier. This paper is structured as follows, we will first go over the difference between classical and quantum software development cycles, then move on to an overview of the tool, its APIs, and a usage example, the results of experimenting with the tool and will be concluded with discussions about related works, current, and future versions of the tool.

\section{Towards a quantum debugging tool}
\label{sec:methodology}

\subsection{The First Step: A Quantum Circuit Slicer}

The current state of classical debuggers is the result of decades of research~\cite{tip1995slicing}, development, and experiments~\cite{kotok1961dec}. One of the most basic concept used in classical debugging is the concept of program slicing~\cite{xu2005brief}. A program tool is used to divide a big body of code into smaller, easy to test and manage chunks. Each of these chunks is called a \emph{slice}. 

Slices are formed in two ways, either manually using breakpoints~\cite{lampson1980processor}, or using a form of automatic/semi-automatic slicing. Using breakpoints, the debugger can divide the code so that the user can observe its behaviour and the variables' contents within each slice. There are different types of automated program slicing, the basic two are static slicing and dynamic slicing. Static slicing works by slicing the program based on a variable or set of variable by eliminating the lines of code that don't include or affect that variable directly or indirectly.  In dynamic slicing, on the other hand, the slice is formed using variable(s) and condition(s). The variable(s) and condition used to form the slices are called the \emph{slicing criteria}.

As discussed in section~\ref{intro}, as the availability, interest, and current size of the available systems continue to increase, the size of circuits implemented will also increase. 
Hence, we implemented a quantum circuit debugging tool that includes a circuit slicer based on the concepts of manual slicing and breakpoints. A quantum circuit slicer will divide a large circuit into smaller, simulatable sub-circuits to prove the use of the circuit both in the current NISQ (Noisy Intermediate-scale Quantum Computers) era~\cite{preskill2018quantum} and the future era of fault tolerant quantum computing.

\begin{figure*}[htbp]
\centerline{\includegraphics[scale=0.7]{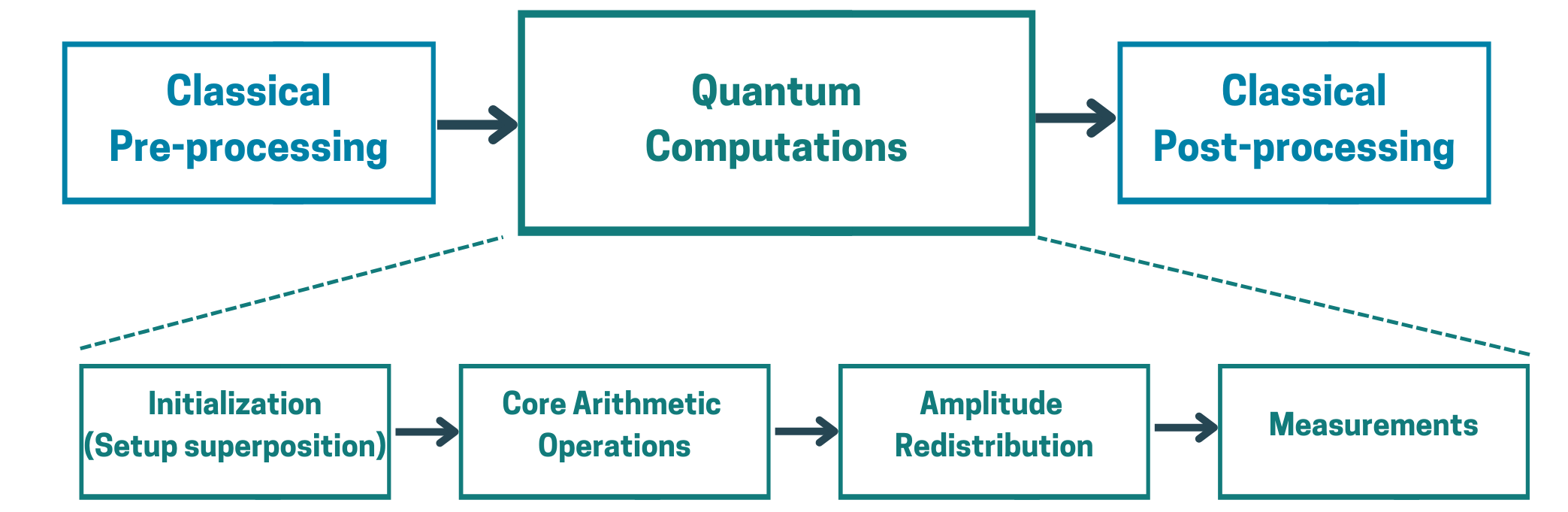}}
\caption{The different steps needed to implement and execute quantum algorithms.}
\label{algex}
\end{figure*}

Generally, quantum algorithms tend to follow a set of steps to solve a problem. All quantum algorithms start with preparing the qubits in a specific state or in a uniform superposition, then perform some arithmetic and calculations, followed by redistribution of the amplitudes. Depending on the algorithm and the problem being solved, some may include some classical pre-processing or post-processing after the measurement procedure Fig.~\ref{alge}.
For example, let us consider Grover's algorithm, which consists of three algorithms steps, preparing the qubits in a uniform superposition, followed by a problem-specific oracle and then a diffusion operator. In the algorithm the oracle and diffusion will repeat multiple times until the answer is reached. 

\begin{figure*}[htbp]
\centerline{\includegraphics[scale=0.3]{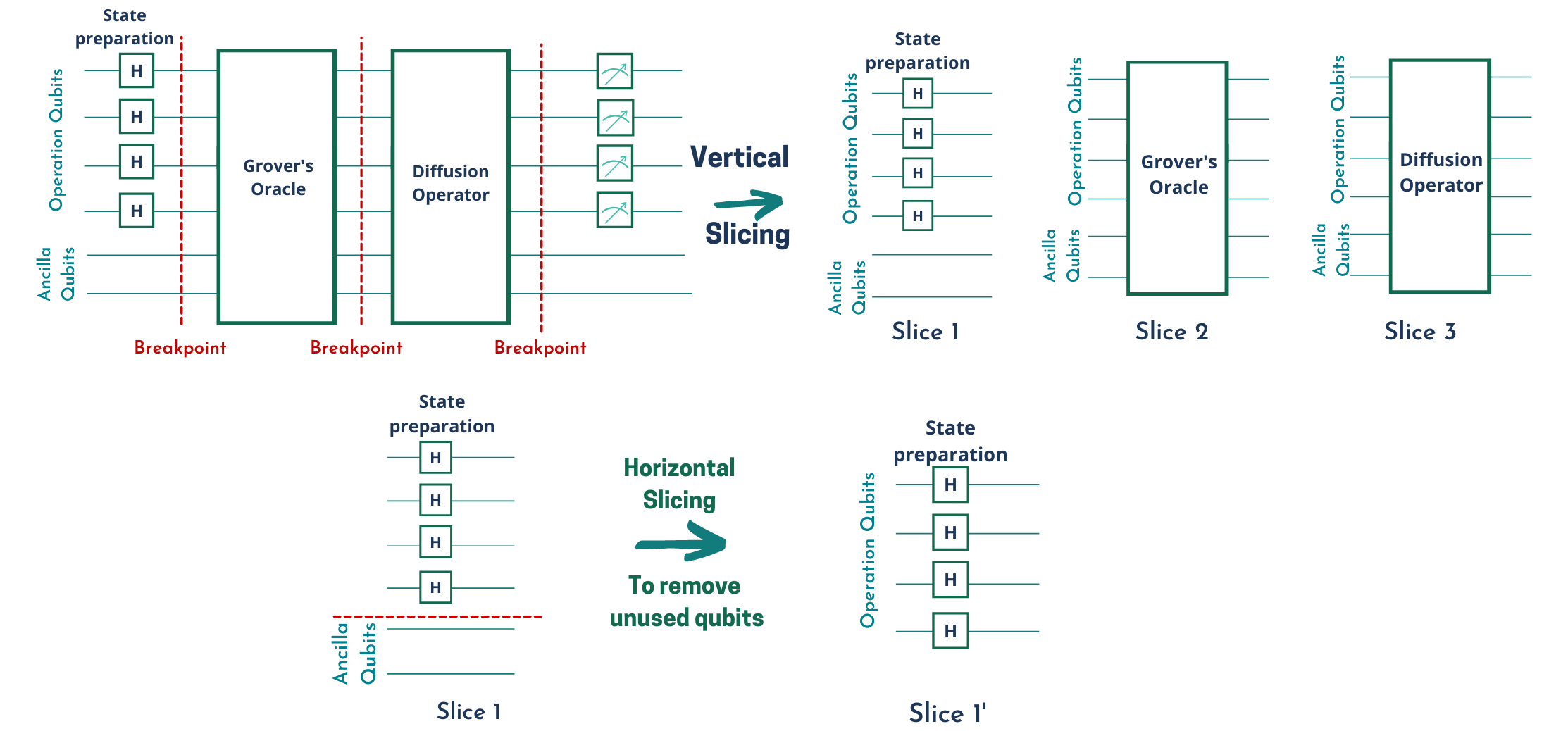}}
\caption{A generic circuit for Grover's algorithm sliced into 3 vertical slices, then the first slice is horizontally re-sliced to remove unused qubits.}
\label{slice}
\end{figure*}

We implemented a manual slicer, where the user inserts breakpoint (in a quantum context, breakbarriers) in the circuit and then simulate the resultant slices or run them on an actual device to observe their behaviour. To make the tool useful for all sizes of circuits and devices, it has to be able to slice the circuits on two axes, the gate axis (vertically) and the register axis (horizontally) Fig.~\ref{slice}. That is, the user can insert breakbarriers vertically in the circuit to divide it into smaller circuits, as well as horizontally to remove any qubits that are unused in any of the slices.

\begin{figure*}[htbp]
\centerline{\includegraphics[scale=0.3]{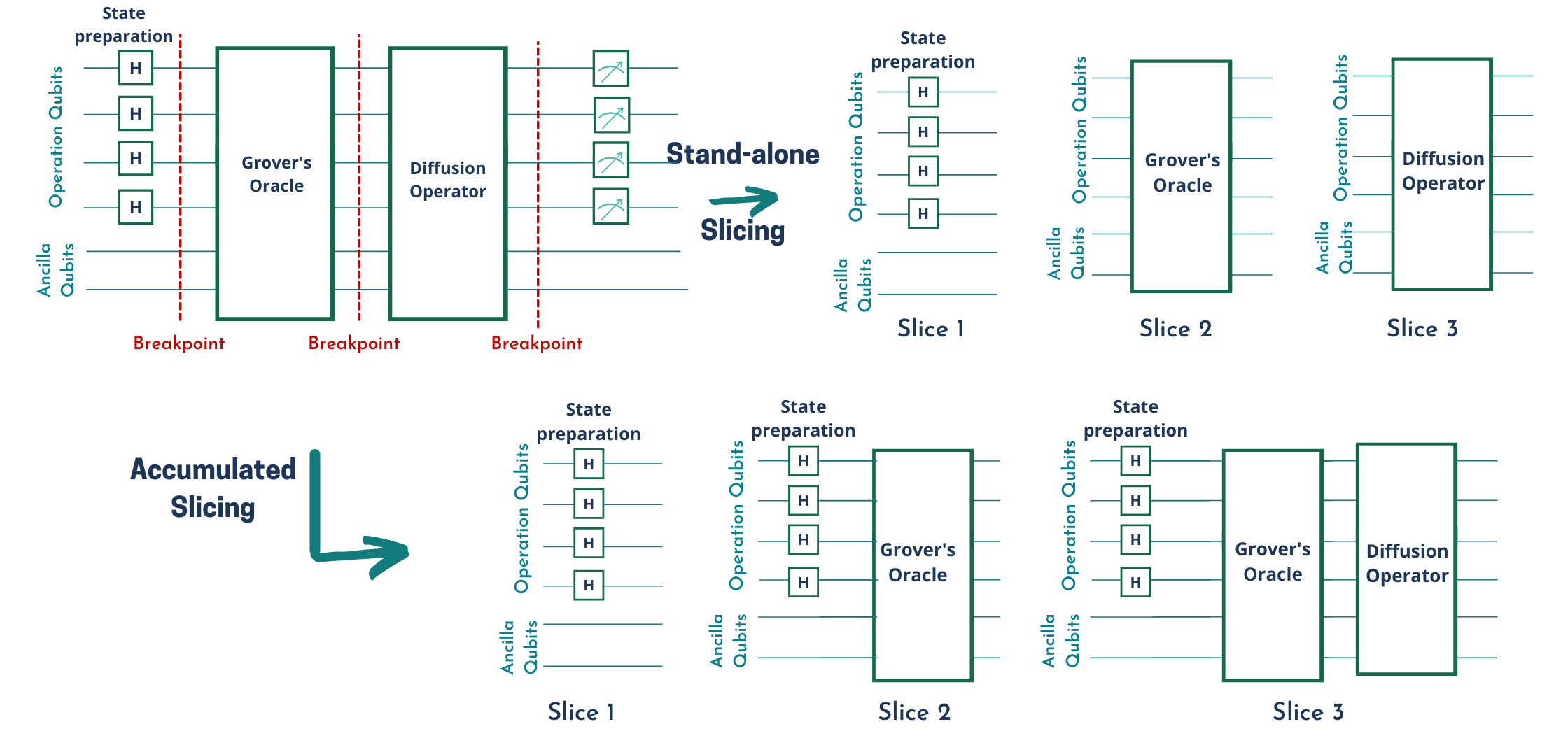}}
\caption{A generic Grover's algorithm circuit sliced using both stand-alone slicing and accumulated slicing.}
\label{slicing}
\end{figure*}

\subsubsection{Vertical Slicing}
To explain the methodology and concept of slicing, let us think of a circuit corresponding to Grover's algorithm~\cite{grover1996fast}. We can use breakbarriers to form divide the circuit into slices based on each algorithmic step. Grover's algorithm consists of three main algorithmic steps: initial state preparation, an oracle, and the diffusion operator. 
In order to keep things simple, assume the Grover's algorithm we are slicing consists of one iteration of the algorithm. to slice this circuit we will insert two breakbarriers, one after the state preparation and one after the oracle. This will result in three sub-circuits each performs a specific step in the overall algorithm. 

The circuit slicer offers two options for vertical slicing Fig.~\ref{slicing}:
\begin{itemize}
  \item Stand-alone slices: the slices are defined by the breakpoints.
  \item Accumulated slices: Each slice is added to the slice before it to create a new slice.
\end{itemize}

\subsubsection{Horizontal Slicing}
Sometimes after slicing the circuit vertically, we may end up with a slice that contains some unused qubits. Since our goal of slicing the circuit is creating smaller, simulatable, executable circuits, having unused qubits is redundant. Hence we can do some horizontal slicing to remove these unused qubits from the slice.
The current version of the tool only allows for automatic slicing of unused qubits. Future expansion will allow users to manually insert horizontal breakbarriers in case of slices with two independent registers or set of qubits.

\subsection{Testing the individual slices}
An important question we need to answer is, \emph{how can we test the individual slices resulted from the tool?} In order to know what debugging or testing technique to use with each slice, we need to be able to obtain the categorization of each slice first.

The current version of the QCS categorizes the slices by investigate the size of the circuit and the type of gates involved in it and how they effect the states entanglement and the interference patterns of the circuit. Moreover, to better help the developer during the testing process, the tool allows the developer to choose a slice and a preparation state to set it in for debugging. The developer can choose to test any given slice on a uniform superposition state, a specific state or a symmetric state. Currently, the tool can prepare any number of qubits in three commonly used symmetric states, the GHZ state, W state~\cite{cruz2019efficient} and Dicke state~\cite{dicke1954coherence}.
In addition to passing the slice and the initial state wanted to the testing function, the user needs to input a test vector (or a list of test vectors) for the tool to examine the slice's behaviour.


\subsection{Gates tracking}
When an error occurs while executing a circuit on an actual hardware, it is often due to one of two reasons, either it is a machine-related one or a semantic (logical) one. Machine-related errors are low-level, hardware-specific faults such as gate errors, readout errors, thermal relaxation errors, measurements errors, etc~\cite{tannu2019mitigating, PhysRevLett.119.180509, PhysRevX.8.031027}. Fixing these types of errors requires implementing quantum error correction, some efficient error minimizing technique~\cite{PhysRevX.7.021050} (or getting very, very lucky on a given run). However, if the circuit is simulated, only semantic errors can occur.

Semantic errors, on the other hand, result in incorrect results on the simulator or even when the machine works properly. These type of errors exist also on the classical realm, and causes the program to misbehave even if the code didn't include syntactic errors.
Semantic errors are more difficult to resolve in quantum circuits than in classical programs, in part because most quantum algorithms select an outcome from a probability distribution at the end of a run, and in part because of their use of superposition, entanglement and interference.  Thus, isolating the section of the circuit containing the error requires careful reasoning and the ability to narrow down operation to make the error as reproducible, and visible, as possible.

One aspect of this is the need for the programmer to work both forward and backward through the tool-chain, examining the circuit at the gate level as well as the higher-level functions that generated the circuit. QCS provides an option for the programmer to track where in the code was each gate added to the circuit. Currently this is done through using an analogous to the traceback information, by printing the line of code, function or module where the gate was added. Another approach we are considering is the including of debugging symbols in an object file, making it easy to match the gate causing the circuit to malfunction to the corresponding line of code.

\subsection{The Tool's API overview}
The tool discussed in this paper is built using Python on top of the Qiskit module. In Qiskit, any quantum circuit is built using an object class QuantumCircuit. Any QuantumCircuit object can contain QuantumRegisters, ClassicalRegisters, different quantum gates and measurement operations. The Qiskit QuantumCircuit object contains many proprieties and characteristics. In order to build our tool, we extended this class to include few new commands to include breakbarriers to cut the circuit and gate tracking option. In addition to these, we added new functionalities to perform the horizontal and vertical slicing. We can divide the functionality the debugging tool adds to Qiskit into two categories, methods added to the QuantumCircuit class and the debugger's core functionality.

\subsubsection{Methods Added to The QuantumCircuit Class}
Since all quantum circuits that can be built using Qiskit use the QuantumCircuit object, we decided to extended that class to include the debugging-needed methods instead of creating a whole new type. That was accomplished by adding two methods:
\begin{enumerate}
  \item {\tt breakbarrier()}: a new object type based on Qiskit's barrier class that is used to pinpoint where the tool is going to cut the circuit when using the Vertical tool function (\textbf{VSlicer}).
  \item {\tt gateInfo()}: a method that, when the debugging mode is enabled, is used to store information about all gates added to the circuit. The information is the gate type, the number of occurrences and where in the code this gate was added to the circuit.
\end{enumerate}

\subsubsection{The Debugger's Core Functionality}
In addition to the methods added to the QuantumCircuit class, we defined new functions that make use of these new methods to enable the debugger functionality. We mainly added four functions.
\begin{enumerate}
  \item {\tt startDebug()}: This function enables the debugging mode by extending the QuantumCircuit Class to include both \emph{breakbarrier} and \emph{gateInfo} methods.
  \item {\tt VSlicer()}: This function takes a QuantumCircuit object that contains breakbarriers and then divides the circuit based on the location of those \emph{breakbarrier} and return the original circuit as well as a list of sub-circuits corresponding to the dividing the circuit based on the \emph{breakbarrier} locations.
  \item {\tt HSlicer()}: This function removes unused qubits or QuantumRegisters from a sub-circuit after using the vertical slicer.
  \item {\tt gateLoc()}:  This function takes a circuit or a sub-circuit and a gate, then displays how many times and where in the code was this gate added to the circuit.
\end{enumerate}


A summary of the methods and functions added to Qiskit to allow the tool to function in Table.~\ref{api}.

\begin{table*}
\centering
\caption{tool API overview}
\label{api}
\resizebox{\textwidth}{!}{
\begin{tabular}{|c|c|c|} 
\hline
\rowcolor[rgb]{0,0.502,0.502} \textcolor{white}{Instruction} & \textcolor{white}{Oprands}                                                                            & \textcolor{white}{Description}                                                                                               \\ 
\hline
\emph{breakbarrier}                                                 & QauntumCircuit method                                                                                 & A method used on a QuantumCircuit object to slice it into smaller sub-circuits.                                              \\ 
\hline
\emph{gateInfo}                                                     & QuantumCircuit method                                                                                 & A method used to keep track of all gates added to the circuit when debugging mode is enabled.                                \\ 
\hline
{\tt VSlicer}                                                     & \begin{tabular}[c]{@{}c@{}}Takes a QuantumCircuit object\\Returns a list of sub-circuits\end{tabular} & Vertical tool function that cuts the circuit based on the breakbarriers added to the QuantumCircuit object.                \\ 
\hline
{\tt HSlicer}                                                      & \begin{tabular}[c]{@{}c@{}}Takes a QuantumCircuit slice\\Returns a list of sub-circuits\end{tabular}  & The horizontal tool function cuts the slice horizontally to remove unused QuantumRegisters.                                \\ 
\hline
{\tt startDebug()}                                                   & Enter debugging mode function                                                                         & Adding this function will start debugging mode by adding \emph{breakbarrier} to the QuantumCircuit object and start gate tracking.  \\ 
\hline
\emph{gateLoc}                                                      & \begin{tabular}[c]{@{}c@{}}Takes a QuantumCircuit and a Gate\\Prints Gate information~\end{tabular}   & A function that displays the information of a certain gate from the \emph{gateInfo} method.                                         \\
\hline
\end{tabular}
}
\end{table*}

\section{The slicer in action}

To better understand how the slicer can help developers understand their circuits and locate errors in it, let's consider an actual example of an implementation of Grover's algorithm applied to the triangle finding problem, where a graph is given and we try to find a 3-node complete graph (\emph{a triangle}) within the larger graph. This problem has been addressed both classically \cite{castellanos2002triangle}, \cite{williams2014finding} and quantumly \cite{le2014improved},\cite{magniez2007quantum}. Although there are various ways this problem can be solved quantumly, using Grover is one of the simplest, most straightforward approaches. 

\begin{figure}[htbp]
\centerline{\includegraphics[scale=0.6]{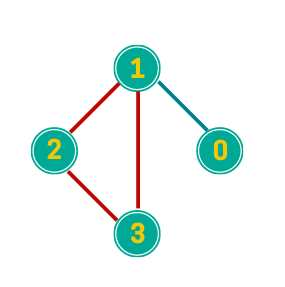}}
\caption{A 4-node graph with a triangle between nodes 1,2, and 3.}
\label{triang}
\end{figure}

For this example problem, let's consider the 4-node graph in Fig.~\ref{triang}. As we mentioned earlier, an arbitrary implementation of Grover's algorithm consist of three algorithmic steps, stage preparation, the oracle and the diffusion operator. The oracle and diffusion operator need to be repeated a total number of {\tt opt\_iter} cyclically. This optimal number of times depends on two factors, the size of the search space $N$ and the number of answers for our problem $m$ (how many triangles in the graph, in this case one) and is calculated using $opt\_iter =  \left\lfloor\frac{\pi}{4}\sqrt{\frac{N}{m}}\right\rfloor $~\cite{boyer1998tight}. In our example, the search space is the entire Hilbert space; which contains {$2^4$} cases $|0000\rangle, |0001\rangle,....., |1111\rangle$. Using that formula and in case of $m = 1$, the {\tt opt\_iter} will be 3.

We will use a simple oracle circuit packaged in a function that aims to mark the correct answer, in our example state $|0111\rangle$.
The code has of three sections, the state preparation, which consist of  Hadamard gates to form the complete Hilbert space, the oracle which marks the state $|0111\rangle$, and the diffusion operator. For this specific example, dividing the circuit into three independent steps was relatively simple, because we just followed the algorithmic steps. 

To implement this circuit we need 6 ancillary qubits, and a 3 qubits a flag that would be in state $|111\rangle$ only if a triangle is found in the graph. We can write this using Python and Qiskit as shown in Listing.~\ref{grov}.\\

\begin{lstlisting}[language=Python, caption=Python and Qiskit code implementing Grover's algorithm for the triangle finding problem., label=grov]
from qiskit import QuantumRegister, ClassicalRegister, QuantumCircuit
import numpy as np    
import math as m
def grover():
    n_nodes = 4
    N = 2**n_nodes  # Hilbert space size 
    #Defone needed qubits
    nodes_qubits = QuantumRegister(n_nodes, name='nodes')
    ancilla = QuantumRegister(6, name = 'anc')
    flag = QuantumRegister(3, name='check_qubits')
    class_bits = ClassicalRegister(n_nodes, name='class_reg')
    tri_flag = ClassicalRegister(3, name='tri_flag')
    qc = QuantumCircuit(nodes_qubits, ancilla, flag, class_bits, tri_flag)
    # Initialize quantum flag qubits in |-> state
    qc.x(flag[2])
    qc.h(flag[2])
    # Initializing i/p qubits in superposition
    qc.h(nodes_qubits)
    # Calculate optimal iteration count
    iterations = round(m.pi/4.sqrt(N))
    #in case of debugging, we will make iteration = 1
    for i in np.arange(iterations):
        oracle(n_nodes, qc, nodes_qubits, ancilla, flag)
        diffusion(qc, nodes_qubits, ancilla)
        qc.breakbarrier() #for debugging only, must be used after startDebug() is called
    return qc
\end{lstlisting}

Now, let's assume we ran this code and the results didn't come out as we expected; that is, the correct answer wasn't marked correctly, which means it doesn't have a higher probability of being measured. To try and locate the error, let's use the slicer to examine each step of the code closely. To do that we will first need to access debugger mode by calling the \textbf{startDebug} function and then apply \textbf{breakbarrier} to the circuit using the \textbf{Vslicer} function with the mini mode.  Applying the \textbf{Vslicer} function to our circuit will result in a list  of three slices. We can rename those slices to make them easier to handle and address as in Listing.~\ref{debug}. 
Since the algorithm will be repeated 3 times, the resultant circuit will end up with repeated, identical parts. Hence, there will be three copies of each the Oracle and the diffusion circuits. So, for debugging purposes, we can focus on one iteration only.\\

\begin{lstlisting}[language=Python, caption=The triangle finding problem in debugging mode., label=debug]
from qcs.qcs import startDebug
from qcs.slicer import Vslicer, Hslicer
#start debug mode of the quantum circuit
QuantumCircuit = startDebug()
qc = grover()
# Vslicer() returns a list; in this example we know a priori it will be length 3 because we have 3 algorithmic steps that we sliced the circuit based on
state_prep, orac, diff = Vslicer(qc, mode = "mini")
#Apply the horizontal slicer on the state preparation and diffusion sections to remove the ancilla and flag qubits so we can only test the operational qubits.
state_prep = Hslicer(state_prep)
diff = Hslicer(diff)
#Testing the state preparation
TestQReg = QuantumRegister(4)
TestCir = QuantumCircuit(TestQReg)
TestCir.append(state_prep, qargs=TestQReg)
\end{lstlisting}

Although the first section in the circuit (the state preparation) is a full-quantum circuit, it is relatively simple to test. If we think about all possible circuits that may fall under the category full-quantum, we see that we can further categorize these circuits. Some which only has a few number of qubits and gates can be tested and verified rather easier than large circuits with many gates. In the case on this example, the circuit only has 4 Hadamard gates applied to 4 qubits.
The Hadamard gate is responsible for creating the superposition. In this case, a superposition over 4 qubits, that is $2^{4}$ states. Testing this slice is not difficult. Before we move on to testing the slice, we first need to remove the qubits that are not used. In the state preparation section of the algorithm, we need to focus on the four qubits representing the nodes of the graph, hence, using the {\tt Hslicer} function we can remove the unused qubits allowing us to examine the four operational qubits.


To test the state prep, we will create a test circuit, and will call it {\tt testCir1}. This circuit will contain 4 qubits and the {\tt state\_prep} slice as in Listing.~\ref{debug}. 

When we run this test, we will get the expected 16 intermediate states which is the equal superposition of all possible 4-qubit states $(\frac{1}{4}\sum_{i=0}^{15}|i\rangle)$. Which leads us to conclude that the error is not in the state preparation section. So, we will move on to the oracle slice.

The oracle slice in this example is a simple pseudo-classical slice, and we can test it by initializing an empty circuit with the same number of qubits as our original circuit and set the flag qubit to the state $|-\rangle$ and then apply the oracle to a superposition of the operational qubits. We can write this in Python as in Listing.~\ref{ortest}.

\begin{lstlisting}[language=Python, caption={Testing the oracle. At the end of this code, {\tt M} contains classical measurement counts that can be compared against expected values, either by hand during debugging or in a scripted test suite for regression and SQA.}, label=ortest]
from qiskit import QuantumRegister, ClassicalRegister, QuantumCircuit, Aer, execute
from qcs.qcs import startDebug
from qcs.slicer import Vslicer, Hslicer
M_simulator = Aer.backends(name='qasm_simulator')
#Testing the oracle
TestQReg = QuantumRegister(4)
Anc = QuantumRegister(6)
flagQubits = QuantumRegister(3)
TestCReg = ClassicalRegister(4)
flagBits = ClassicalRegister(3)
TestCir = QuantumCircuit(TestQReg, TestCReg, Anc, flagQubits, flagBits)
TestCir.append(orac, qargs=[TestQReg, Anc, flagQubits])
#Measuring the operation and flag qubits
qc.measure(TestQReg,TestCReg)
qc.measure(flagQubits, flagBits)
M = execute(qc, M_simulator, shots= 1024).result().get_counts(qc) 
\end{lstlisting}

Running this test will result in multiple states with different combination of the node qubits and the flag qubits with different possibilities based on the number of shots used in the \emph{execute} function. For example, a run to the oracle testing slice will get a result like, $(26|0101\rangle|000\rangle,..., 33|0111\rangle |111\rangle, 35|0110\rangle|000\rangle)$. 
The numbers before each state is the probability of the state being measured within 1024 shots. These numbers are not constant and will differ on each run and with different number of shots. What we are looking for in this test is a case where the flag qubits are in the state $|111\rangle$, the operational qubits state that accompany this flag qubits state is the answer the oracle's marks. In this test, we can see that the oracle marks the correct answer which is $|0111\rangle$. In our example, the number of states we need to examine is small and can be tracked by eye, which is not the case as the circuit size increases. In those cases, a simple program can be written to search for the $|111\rangle$ flag qubits state, using regular expression or another search technique that will make the process of finding what the oracle marks faster and more efficient.\\

\begin{figure}[htbp]
\centerline{\includegraphics[scale=0.28]{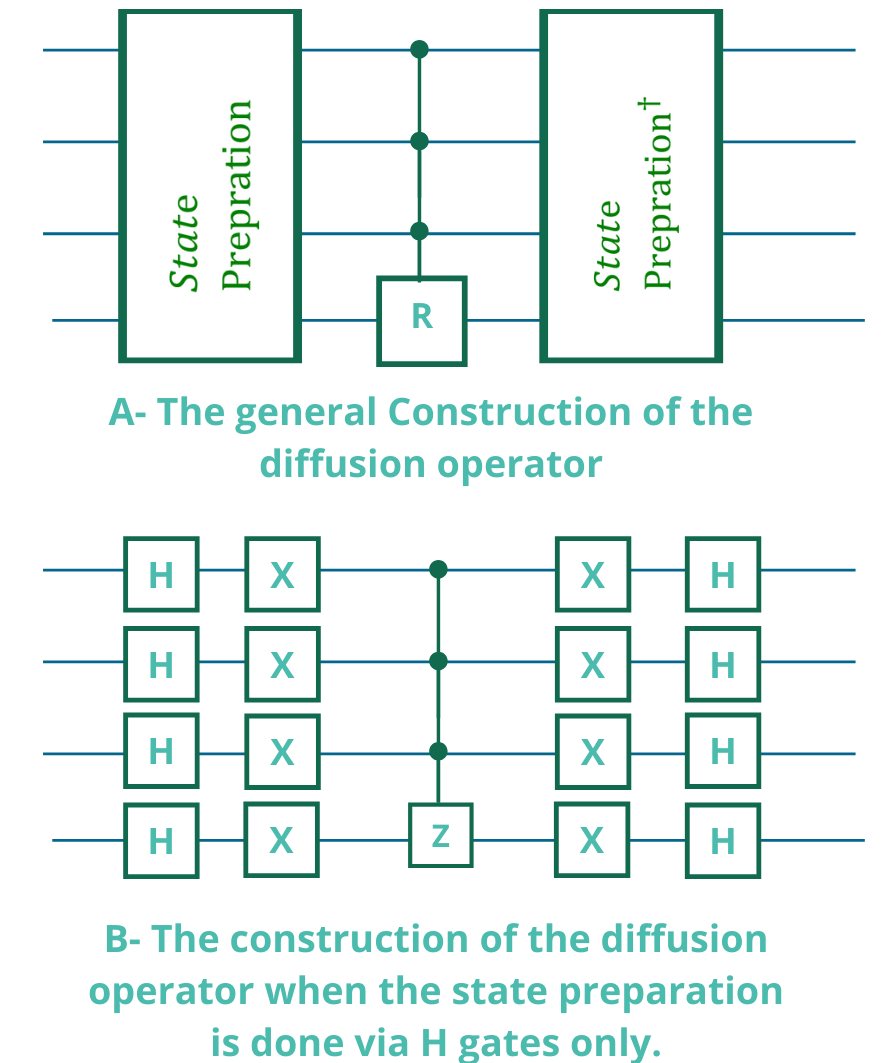}}
\caption{The general construction of the diffusion operator.}
\label{diffusion}
\end{figure}

The last part of the circuit we need to test is the diffusion operator. Since we already tested the other parts of the circuit and made sure that they function properly, we can, with some confidence, say that the error occurs in the diffusion operator. The diffusion operator is added to the circuit to execute a rotation about the average to increase the probability of measuring the correct answer, or the answer marked by the oracle. Generally, if we used the Hadamard gates as the form of state preparation, the diffusion operator should look as shown in Fig.~\ref{diffusion}-A. 
For our example, the diffusion operator code is shown in Listing.~\ref{diff}

\begin{lstlisting}[language=Python, caption=The function constructing the diffusion operator (that results in an error due to an extra NOT gate in line 7)., label=diff]
def grover_diff(qc, nodes_qubits, ancilla):
    qc.h(nodes_qubits)
    qc.x(nodes_qubits)
    #Apply 3 control qubits Z gate
    cnz(qc,len(nodes_qubits)-1,nodes_qubits,ancilla)
    qc.x(nodes_qubits)
    qc.x(nodes_qubits[0])
    qc.h(nodes_qubits)
\end{lstlisting}

Now we know that the reason our circuit is giving a wrong answer is probably because of a bug in the diffusion operator (a full-quantum circuit). Unlike the state preparation, the diffusion operator is somewhat more difficult to test. So, we need to look at the definition of the diffusion operator. Mathematically, the diffusion operator D is defined as D = $state\_prep$ R $state\_prep^{\dagger}$, where $R$ is a a zero reflection or a zero-phase shift. This phase shift can be calculated by $R = 2|0\rangle^{\otimes n}\langle 0|^{\otimes n}- I_{n}$, where $I _{n}$ is the identity matrix on n qubits~\cite{grover1996fast, brassard2002quantum, nielsen2002quantum}. When we examine this equation and try to test if the function in Listing.~\ref{diff} correctly implement it, we can see that we have something missing. 

We can tell that our diffusion function is missing some NOT gates. To correctly implement the diffusion operator, we need a multi-controlled Z gate sandwiched by NOT gates. Because we have 4 qubits, then we will need a 3\-control Z gate surrounded by 8 NOT gates Fig.~\ref{diffusion}-B. 

Now, we can use the \emph{gateLoc} function to obtain extra information about the NOT gate within the diffusion operator. We can simply do that by calling the function as gateLoc(diff, "x").

\begin{lstlisting}[language=bash, caption=The output of the gateLoc function when querying the NOT gate in the diffusion operator, label=diffgate]
--------------------------
There are 3 times where the x gate was added to the circuit.
It was added to the circuit in the following locations:
File_directory, File_name, line 25, in <module>
    diffusion(qc, nodes_qubits, ancilla)

File "file_name", line 3, in grover_diff
    qc.x(nodes_qubits)
    
File "file_name", line 6, in grover_diff
    qc.x(nodes_qubits)
    
File "file_name", line 7, in grover_diff
    qc.x(nodes_qubits[0])
--------------------------
\end{lstlisting}

Examining the output of \emph{gateLoc} in Listing.~\ref{diffgate}, we can see that the slicer tells us that the NOT gate were added to the circuit 3 times. We should pay attention here that two of those three times, the gate was applied to an entire register. That is, in lines 3 and 6 of the diffusion function Listing.~\ref{diff} that NOT was applied to all qubits within the {\tt nodes\_qubits} register. So, based on that we can conclude that the NOT gate was added a total of 9 times and not the correct amount of 8. By removing the extra NOT gate added in line 7 in Listing.~\ref{diff}, and then testing the entire circuit, we can see that we successfully located the error and fixed it.

The example in this section is a simple, short one for the purpose of illustrating the functionality of the debugger. However, the same concepts and methodology applied to it can be extended to any other quantum circuit with any size and number of qubits.
\section{Evaluation}
\label{sec:eval}
Our main goal in developing a quantum circuit slicing tool is to assist developers and quantum computing learners to better understand their circuits and to be able to extend and alter them. In order to evaluate our tool, we released a beta version to a group of quantum programmers/ researchers with different levels of expertise in the field of quantum computing, from absolute beginners to experts.

We focused on assessing a couple factors: the ease of installing and using the tool, the usefulness of the tool in understanding the circuit and how much it helped in locating a bug in exercises we created or a quantum circuit containing an error that the programmers/ researchers chose. In addition, we sought out how the programmers/ researchers utilized the tool in order to decide the development of future versions of the tool. We as well as a group of researchers used the beta version of the slicer on some quantum circuits and answer a survey about their experiences using the tool and the properties of the circuits they are used it on. The algorithm names, size, and other information about the circuits used to test the slicer are in Table~\ref{exps}. All of the circuits used for the initial testing of the tool contained reproducible bugs resulting from either an incorrect implementation of the algorithm or by an error in applying the gates to certain qubits. 

\begin{table*}[]
\centering
\caption{QCS experiments details displaying information about the circuits used in experiments based on Qiskit}
\label{exps}
\resizebox{\textwidth}{!}{
\begin{tabular}{|c|c|c|c|c|c|c|c|c|} 
\hline
\rowcolor[rgb]{0,0.502,0.502} \textcolor{white}{Algorithm} & \textcolor{white}{LOC} & \textcolor{white}{Circuit size} & \textcolor{white}{No. of qubits} & \textcolor{white}{gates} & \textcolor{white}{No. of slices} & \textcolor{white}{No. of algorithmic steps} & \textcolor{white}{Ave. size of the slice}  \\ 
\hline
{\cellcolor[rgb]{0.612,0.878,0.839}}QFT                    & 8                      & 7                               & 3                                & [3*H, 3*CP, 1*SWAP]                        & 4                                & 5                                           & 2 gates/ slice                                                      \\ 
\hline
{\cellcolor[rgb]{0.612,0.878,0.839}}Full Adder                 & 26                     & 11                               & 4                                &  [3*CCNOT, 5*CNOT]                      & 4                                                                           & 4                                         & 2 gates\/ slice     \\ 
\hline

{\cellcolor[rgb]{0.612,0.878,0.839}}Simon's algorithm for 2 qubits with the secret string 11                    & 28                      &11                                & 6                                & [6*H, 5*CNOT]                        & 3                                & 3                                           & 3 gates/ slice                                                      \\ 
\hline
{\cellcolor[rgb]{0.612,0.878,0.839}} Grover for triangle finding                      & 115                      & 172                               & 13                                & [42*CCNOT, 25*H, 17*X, 16*CNOT, 12*MCCNOT, 2*CZ]                      & 3                                & 3                                           & 28 gates/ slice                         \\
\hline
{\cellcolor[rgb]{0.612,0.878,0.839}} 4-qubit Quantum counting                     &86                      & 470                               & 8                               & [192*H, 180*NOT, 30*CCNOT, 30*MCT, 30*Z, 2*SWAP, 6*CU1]                      & 6                                & 3                                          & 40 gates/slice                         \\
\hline
\end{tabular}
}
\end{table*}

After using the current version of the tool, the software developers  agreed that using the tool made locating the bug in the circuit a more efficient task. Using the tool helped them decrease much of the manual load they would often do when attempting to debug and test their quantum circuits.

\section{Related Work and Discussion}
\label{dis}

Debugging quantum computing has been a concern of many researchers since the beginning or quantum computing. Some researchers have been making more progress designing and establishing a framework to debug computing in the recent years, targeting different levels of the design. 
If we desire to debug a quantum computer, we will need to be more specific about the level of implementation the debugger will be targeting. In quantum computing, there are different levels of implementation starting with the higher level, which is the algorithmic implementation, then the circuit level, and the lower level the consists of electrical pulses executed on an actual hardware.
One of the examples of a quantum debugger is the effort done by Q\#. Q\# is a full-on quantum development environment developed by Microsoft to be similar to C\#. Q\# offers various ways to debug a quantum circuit, in particular NISQ era quantum programs. They offer the use of one of the most common testing approaches in classical debugging, which is unit testing. Unit testing is a useful testing and debugging technique if the results of the circuit are known. If the circuit behaviour is not easy to understand however, writing test cases for it will not be an easy task and hence unit testing will not be as useful. 
Another tool Q\# offers, is facts and assertion  and those deal in case your function or circuit does not run on the hardware or if the hardware returns a "no results".

Besides Q\#, there is some work done on debugging quantum circuits at run time~\cite{Li}, or using statistical methods~\cite{huang2019statistical}. Some work have also been done on trying to establish a quantum programming development cycle, similar to the classical software cycle to suggest different techniques that can be used in each step~\cite{weder2020quantum} \cite{miranskyy2021testing} and the debugging tactics for quantum algorithms~\cite{miranskyy2021testing}.

Moreover, since understanding the flow of quantum programs and the causes of errors is essential to the ability of debugging quantum circuits, researchers focused on reproducible bugs and categorizing the occurrences of bugs in quantum programs~\cite{zhao2021bugs4q,campos2021qbugs}. In one of these works~\cite{luo2022comprehensive}, the possible bugs that might occur when implementing quantum algorithms were divided into 4 categories: bugs due to the API used to implement the algorithm, bugs due to a fault in the logic of the application, math-related bugs, or bugs due to an error aside of the above reasons.

All these efforts focus on addressing the debugging of quantum programs, none actually offers a systematic approach to debugging quantum circuits. The tool proposed in this paper should help the developers find and fix the different types of bugs from one caused because of the API used to the ones resultant from a faulty implementation of the algorithm. In addition, we address the higher level algorithmic side of the implementation. Our tool does not only aim at debugging a circuit, but to also help programmers understand their circuit better. Because if they do, they will be able to perform more efficient debugging process.
\section{Conclusion \& Future Work}
\label{conc}

When dealing with quantum circuits, validating the correctness of the answer remains the biggest challenge quantum technology faces today and if a debugging tool is not developed, this challenge will get more difficult as the circuit size increase and the ability of current hardware improves. 

We tested the functionality of the debugging tool by releasing a beta version for a small group of developers to test and use on some small and medium-scale circuits. The developers used the core functionality of the tool to simplify their circuits, better understand its behaviour, and locate bugs using the gate tracker functionality. After using the tool, the developers answered a survey to help us understand their experience and assist us in improving the tool in the future. Moreover, we examined how our tool would fit in with other existing efforts towards better understanding quantum circuits and hence debugging them.

The debugging tool proposed in this paper is the start of a larger tool that can transform the future of debugging and understanding quantum circuits and we plan on expanding it further in the future. Further expansion of the tool that are currently underdevelopment is:

\begin{itemize}
    \item Tune the categorization criteria to more accurately categorize the slices. We believe a more detailed categorization will lead to easier debugging.
    \item Expand the gate tracking function to make it more flexible and to make the query function more efficient by letting the user define further information about the gate in question, such as specific range in the code or within a function.
    \item Develop an automated testing function for commonly used subroutines in quantum algorithms such as QFT, quantum walks and Grover's diffusion operator.
\end{itemize}

Finally, to achieve our optimal goal of easing up the process of building, understanding and debugging quantum circuits, we are working on developing a graphical interface where the user can slice, examine and test the circuit without needing to write any code.


\section*{Acknowledgment}
This work was supported by MEXT Quantum Leap Flagship Program Grant 
Number JPMXS0118067285.

\bibliographystyle{./IEEEtran}
\bibliography{bib}

\end{document}